\documentclass[journal]{IEEEtran}

\ifCLASSINFOpdf
\else
   \usepackage[dvips]{graphicx}
\fi
\usepackage{url}

\usepackage{graphicx}
\usepackage[colorlinks,allcolors=blue]{hyperref}
\usepackage{amsthm}
\usepackage{mathtools}
\usepackage{amssymb}
\usepackage{cite}

\usepackage{pgfplots}
\usepackage{tikz}
\usetikzlibrary{matrix}
\usetikzlibrary{arrows.meta} 
\usepackage[noabbrev]{cleveref}

\usepackage{mathtools,bbm}
\newtheorem{definition}{Definition}
\newtheorem{remark}{Remark}
\begin{document}

\title{Parallel state estimation for systems with integrated measurements}

\author{Fatemeh Yaghoobi and Simo Särkkä, \IEEEmembership{Senior Member, IEEE}
\thanks{This work was funded by the Research Council of Finland (project 321900).}
\thanks{F. Yaghoobi and S. Särkkä are with the Department of Electrical Engineering
and Automation, Aalto University, 02150 Espoo, Finland (email:
fatemeh.yaghoobi@aalto.fi, simo.sarkka@aalto.fi).}}

\markboth{Journal of \LaTeX\ Class Files, Vol. 14, No. 8, August 2015}
{Shell \MakeLowercase{\textit{et al.}}: Bare Demo of IEEEtran.cls for IEEE Journals}
\maketitle

\begin{abstract}
    This paper presents parallel-in-time state estimation methods for systems with Slow-Rate inTegrated Measurements (SRTM). Integrated measurements are common in various applications, and they appear in analysis of data resulting from processes that require material collection or integration over the sampling period. Current state estimation methods for SRTM are inherently sequential, preventing temporal parallelization in their standard form. This paper proposes parallel Bayesian filters and smoothers for linear Gaussian SRTM models. For that purpose, we develop a novel smoother for SRTM models and develop parallel-in-time filters and smoother for them using an associative scan-based parallel formulation. Empirical experiments ran on a GPU demonstrate the superior time complexity of the proposed methods over traditional sequential approaches.
\end{abstract}

\IEEEpeerreviewmaketitle

\section{Introduction}
\label{sec:intro}
    \IEEEPARstart{T}{his} paper aims to develop parallel-in-time state estimation methods for systems with integrated measurements. Such state estimation problems arise in various industrial and laboratory scenarios where the measurements are formed as integrals or sums of states over the sampling period \cite{fatehi2017state,salehi2018state,yaghoobi2020bayesian}. In particular, we consider Slow-Rate inTegrated Measurement (SRTM) scenarios, where the measurements are obtained at a relatively slow rate, and we are interested in inferring the states not only at the slow measurement rate but also at a faster rate over the sampling period. 
    

    In the context of state estimation problems, several studies have explored integrated measurements. For linear discrete-time state-space models (SSMs) in the presence of SRTMs, Fatehi et al. \cite{fatehi2017state} proposed a modified Integrated Measurement Kalman Filter (IMKF) formulation for fast-rate and slow-rate state estimation, while Salehi et al. \cite{salehi2018state} investigated uncertainty in process and measurement noises in such scenarios. For nonlinear SRTMs, Yaghoobi et al. \cite{yaghoobi2020bayesian} developed Bayesian filtering formulations for general nonlinear non-Gaussian SSMs and solved the problem using a sequential Monte Carlo method. Han et al. \cite{han2020state} studied multirate measurement nonlinear systems in the presence of integral terms and variable delays. Guo et al. \cite{guo2015state} discussed the state estimation problem in the presence of integrated, infrequent, and delayed measurements, proposing an unscented Kalman filter (UKF) method to estimate the states. Other studies have also investigated parameter estimation and control problems with SRTMs \cite{kheirandish2020identification, salehi2022robust, salehi2021offline, khosro2020design}.
    
    All the mentioned Bayesian filtering methods are inherently sequential, with a linear time complexity in the number of time steps. To address the computational time associated with Kalman filters and smoothers, Särkkä et al. \cite{sarkka2020temporal} introduced parallel-in-time formulations of Bayesian filters and smoothers, which deliver identical state estimation results while reducing the linear time complexity to logarithmic. These methods have also been extended in various ways \cite{hassan2021temporal,yaghoobi2021parallel,yaghoobi2022parallel}. However, these methods cannot be applied to SRTMs as such, because the state estimation methods for SRTMs differ from standard Kalman filters and smoothers.

    The contribution of this paper is as follows. For linear SSM with integrated measurement, we derive a novel smoothing solution for both slow-rate and fast-rate states. We then introduce parallel formulations for filtering and smoothing methods in the SRTM model, reducing the linear time complexity of state estimation to logarithmic (span) time complexity. We experimentally evaluate the proposed algorithms on GPU using the JAX framework. In the following,  Section~\ref{sec:background-parallel} reviews the parallel framework for Bayesian filters and smoothers. Sections~\ref{sec:problem-formulation} and \ref{sec:proposed-method} present the problem formulation and our method. Section~\ref{sec:experiment} assesses methods via a numerical experiment, and Section~\ref{sec:conclusion} concludes the paper.
\section{Parallel Framework for Conventional Bayesian Filters and Smoothers}
\label{sec:background-parallel}
    Given a state-space model of the following form:
    \begin{equation} \label{eq:state-space-model}
      x_k \mid x_{k-1} \sim p(x_k \mid x_{k-1}), \quad y_k \mid x_k \sim p(y_k \mid x_k),
    \end{equation}
    the goal of Bayesian filtering is to find the filtering distributions $\big(p(x_k\mid y_{1:k})\big)|_{k=1}^N$, where $N$ is the total number of time steps, whereas a Bayesian smoother aims to compute the smoothing distributions given all data, that is, $\big(p(x_k \mid y_{1:N})\big)|_{k=1}^N$. The filtering and smoothing distributions can be computed with the classical sequential Bayesian filters and smoothers \cite{sarkka2023bayesian}. 
    In the case of linear SSM with Gaussian noise, they reduce to the closed-form Kalman filter~\cite{kalman1960new} and Rauch-Tung-Striebel (RTS) smoother~\cite{rauch1965maximum}. These sequential algorithms have a linear $O(N)$ execution time \cite{sarkka2023bayesian,sarkka2020temporal}. 
    
    To speed up the computations of the mentioned methods on GPU-based systems, the authors of \cite{sarkka2020temporal} proposed a parallel-in-time framework for Bayesian filters and smoothers based on the parallel-scan algorithm. The parallel-scan algorithm~\cite{blelloch1989scans} is a general method computing prefix-sums of sets of elements $\{a_k\}|_{k = 1}^N$ with respect to a binary associative operator $\otimes$: $(a_1, a_1 \otimes a_2, \ldots, a_1 \otimes \cdots \otimes a_N)$. The method reduces the time complexity from $O(N)$ to $O(\log N)$ via the use of parallelization. This resulting algorithm  \cite{sarkka2020temporal} is the following.
    \begin{definition}
    Given two positive functions $g_i(y)$ and $g_j(y)$, and two conditional densities $f_i(x \mid y)$ and $f_j(x \mid y)$, we define the binary operator $(f_i,g_i) \otimes (f_j,g_j) \coloneqq (f_{ij},g_{ij})$ as
    \begin{equation}
    \begin{split}
        f_{ij}(x \mid z) = \frac{\int{g_j(y)f_j(x \mid y)}f_i(y\mid z)dy}{\int{g_j(y)}f_i(y \mid z)dy},\\
        g_{ij}(z)=g_i(z)\int{g_j(y)f_i(y \mid z)dy}.
    \end{split}
    \label{eq:gen-filter-op}
    \end{equation}
    Under this parameterization, \cite[Theorem 3]{sarkka2020temporal} states that $\otimes$ is associative and, by selecting elements $a_k = (f_k, g_k)$ as
    \begin{equation}  \label{a_k_filter-app}
        \begin{split}
            f_k(x_k \mid x_{k-1}) &= p(x_k \mid y_k, x_{k-1}), \\
            g_k(x_{k-1}) &= p(y_k \mid x_{k-1}),
        \end{split}
    \end{equation}
    where $p(x_1 \mid y_1,x_0) = p(x_1 \mid y_1)$ and $p(y_1 \mid x_0)= p(y_1)$, the results of the $k$-th prefix sum of the elements $a_j$ under $\otimes$ recovers the filtering distribution at step $k$ as well as the marginal likelihood of the observations, $a_1\otimes \dots \otimes a_k = \binom{p(x_k \mid y_{1:k})}{p(y_{1:k})}$.
    \end{definition}
    After obtaining the parallel filtering results, in order to derive parallel smoothing results, new element $a_k$ and the binary associative operator $\otimes$ are defined \cite{sarkka2020temporal} as follows.
    
    \begin{definition}
        For any conditional densities $f_i(x \mid y)$ and $f_j(x \mid y)$, the binary operation $f_i \otimes f_j \coloneqq \int{f_i(x|y)f_j(y|z)}dy,$ is associative and by selecting 
        \begin{equation} \label{eq:a-k-smoother}
            a_k = p(x_k \mid y_{1:k}, x_{k+1}),
        \end{equation}
    with $a_N=p(x_N \mid y_{1:N})$, the Bayesian smoothing solutions can then be calculated as $p(x_k \mid y_{1:N}) = a_k \otimes a_{k+1} \otimes \dots \otimes a_N $.
    \end{definition}
    Because the operators $\otimes$ of both the filters and smoothers are associative, the parallel-scan algorithm~\cite{blelloch1989scans} can be used to parallelize the computations. Furthermore, for linear Gaussian models the above operators and elements have closed-form representations \cite{sarkka2020temporal}.
    In this paper, we aim to derive the elements $a_k$ and the binary associative operator $\otimes$ for linear Gaussian SRTM systems, giving parallelizable formulations for both filtering and smoothing solutions for them.
\section{Problem formulation}
\label{sec:problem-formulation}
    Let us consider the linear discrete-time SSM with additive Gaussian noise in the presence of SRTMs which has the form:
    \begin{subequations} \label{eq:original-linear-SSM}
        \begin{align}
            \begin{split} \label{eq:x-t}
                x_{t+1} &= A x_t + B u_t + w_t, \quad \forall t
            \end{split}\\
            \begin{split} \label{eq:y_tl}
                y_{t_j} &=  \frac{C}{l}\, \sum_{i=1}^l\, x_{t_l-l+i} + v_{t_j}, \quad t_j = t_{j - 1} + l.
            \end{split}
        \end{align}
    \end{subequations}
    The variable $t$ represents the fast-rate sampling time, $t_j$ denotes the slow-rate sampling time with $j = \{1, \ldots, N\}$, and $t_1 = l$. Furthermore, $x_t \in  \mathbb{R}^{n_x}$ is the fast-rate state, and $y_{t_j} \in \mathbb{R}^{n_y}$ is the slow-rate integrated measurement updated every $l$ samples. In the model, $A$ is the state transition matrix, $B$ is the input matrix, and $C$ is the measurement matrix which relates the average of $l$ fast-rate states on the $j$th interval to the integrated measurement $y_{t_j}$. The Gaussian initial state, process and measurement noises are $x_0 \sim \mathcal{N}(x_0; m_0, P_0)$, $w_t \sim \mathcal{N}(w_t; 0, Q)$ and $v_{t_j} \sim \mathcal{N}(v_{t_j}; 0, R)$, respectively. Finally, $u_t \in  \mathbb{R}^{n_u}$ is the input of the system.

\begin{figure}[t]
        \centering
        \resizebox{0.8\columnwidth}{!}{\begin{tikzpicture}
\matrix[matrix of math nodes,column sep=2em,row
sep=1em,cells={nodes={circle,draw,minimum width=1em,inner sep=1pt}},
row 1 column 2/.style={nodes={circle,draw}},
row 1 column 3/.style={nodes={circle,draw}},
row 1 column 5/.style={nodes={circle,draw}},
row 1 column 1/.style={nodes={circle,draw=none}},
row 1 column 4/.style={nodes={circle,draw=none}},
row 1 column 6/.style={nodes={circle,draw=none}}] (m) {
  \ldots &   x_{k, 1}  & x_{k, 2}  &   \ldots &  x_{k, l} & \ldots \\
  & & & & y_k & \\
};
\draw[->] (m-1-1) -- (m-1-2);
\draw[->] (m-1-2) -- (m-1-3);
\draw[->] (m-1-3) -- (m-1-4);
\draw[->] (m-1-4) -- (m-1-5);
\draw[->] (m-1-5) -- (m-1-6);
\draw[->] ([yshift=-0.6ex]m-1-5.south-|m-1-5.south) -- (m-2-5.north);
\draw[decorate,decoration={brace,raise=-0.8mm,amplitude=5pt}]([yshift=0ex]m.east-|m-1-5.east) -- ([yshift=0ex]m.west-|m-1-2.west);
\end{tikzpicture}}
        \caption{Graphical representation of one interval of SSM in a SRTM system.}
        \label{fig:integrated_model}
    \end{figure}
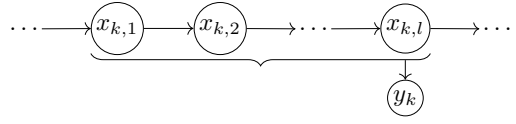

    To have a better representation of fast-rate and slow-rate states, the dual-rate SSM is reformulated as follows \cite{salehi2018state}:\footnote{It is worth noting that an alternative approach would be to augment the integrated state to the state, but here we adopt the approach compatible with \cite{salehi2018state} for the sake of computational efficiency.}
    \begin{subequations} \label{eq:dual-rate-ssm1}
        \begin{align}
            \begin{split} \label{eq:x-k-i}
                x_{k, l - i + 1} &= A x_{k, l - i} +B u_{k, l - i} + w_{k, l - i},
                \end{split}\\
                \begin{split} \label{eq:y-k-l}
                y_{k} &= \frac{C}{l}\sum\limits_{i=1}^{l} x_{k, l - i + 1} + v_{k},
            \end{split}
        \end{align}
    \end{subequations}
    where $k = \{1, \ldots, N\}$ represents the index of slow-rate samples, while $i = \{1, \ldots, l\}$ indexes the fast-rate samples within each slow-rate interval. We also denote $\{k, 0\} = \{k-1, l\}$ and $x_{0, l} = x_{0}$. As the measurements are available only at slow-rate sample times, a single index is used to represent them. Figure~\ref{fig:integrated_model} illustrates one interval of the SSM in a SRTM system.

    \subsection{Sequential IMKF method for slow-rate states}
    Here, we aim to obtain filtering results for slow-rate states, that is, $\big(p(x_{k, l} \mid y_{1:k})\big)|_{k=1}^N$. We first integrate out the fast-rate states $x_{k, 1:l-1}$, which leads to a dynamic model \cite{salehi2018state}
    \begin{equation} \label{eq:h-k-l-2}
        x_{k, l} = \Bar{A} x_{k-1, l} + \Bar{B} \, \Bar{u}_k + \Bar{G} \, \Bar{w}_{k},
    \end{equation}
    where $\Bar{A} = A^l$, $\Bar{G} = \begin{bmatrix}
                A^{l-1} & A^{l-2} & \cdots & I
            \end{bmatrix},$ and
    \begin{equation}
        \begin{split}
            \Bar{B} &= \begin{bmatrix}
                A^{l-1} B & A^{l-2} B & \cdots & B
            \end{bmatrix},\\
            \Bar{u}_k &= \begin{bmatrix}
                u_{k-1, l} & u_{k, 1} & \cdots & u_{k, l-1}
            \end{bmatrix}^\top,\\
            \Bar{w}_k &= \begin{bmatrix}
                w_{k-1, l} & w_{k, 1} & \cdots & w_{k, l-1}
            \end{bmatrix}^\top.
        \end{split}
    \end{equation}
    We assume that the previous filtering results, $p(x_{k-1, l} \mid y_{1:k-1}) \sim \mathcal{N}(x_{k-1, l}; m^f_{k-1, l}, P^f_{k-1, l})$, with $(m^f_{0, l}, P^f_{0, l}) = (m_0, P_0)$, are available. We then get the following.
    
    \textbf{Prediction of the slow-rate state, that is, computation of $p(x_{k, l} \mid y_{1:k-1}) \sim \mathcal{N}(x_{k, l}; m_{k, l}^-, P_{k, l}^-)$} can be done using \eqref{eq:h-k-l-2}:
    \begin{equation} \label{eq:prediction}
        \begin{split}
            m_{k, l}^- &= \Bar{A} m^f_{k-1, l} + \Bar{B} \, \Bar{u}_k,\\
            P_{k, l}^- &= \Bar{A}P^f_{k-1, l} \Bar{A}^T + \Bar{Q},
        \end{split}
    \end{equation}
    where $\Bar{Q} = \Bar{G} \, \Tilde{Q} \,  \Bar{G}^\top$, and $\Tilde{Q} = \mathbb{E}[\Bar{w}_k\Bar{w}_k^\top]$.
    
    \textbf{Update of the slow-rate state, that is, computation of $p(x_{k, l} \mid y_{1:k}) \sim \mathcal{N}(x_{k, l}; m^f_{k, l}, P^f_{k,l})$} can be done by reformulating the measurement model in Equation~\eqref{eq:y-k-l} as a function of $x_{k-1, l}$ as follows \cite{salehi2018state}:
    \begin{equation} \label{eq:y-k-l-2}
        y_{k} = \Bar{C} x_{k-1, l} + \Bar{D} \, \Bar{u}_k + \Bar{M} \, \Bar{w}_{k} + v_{k},
    \end{equation}
    where $\Bar{C} = \frac{C}{l} \sum_{i=1}^l A^{i},$ and
    \begin{equation}
        \begin{split}
            \Bar{M} &= \frac{C}{l} \, \begin{bmatrix}
                \sum\limits_{i=0}^{l-1} A^{i} & \sum\limits_{i=0}^{l-2} A^{i} & \cdots & I
            \end{bmatrix},\\
            \Bar{D} &= \frac{C}{l} \, \begin{bmatrix}
                \sum\limits_{i=0}^{l-1} A^{i} B & \sum\limits_{i=0}^{l-2} A^{i} B & \cdots & B
            \end{bmatrix}.
        \end{split}
    \end{equation}
    Then, the resulting update follows from the Gaussian distribution conditioning rule \cite[Theorem 6.9]{sarkka2023bayesian} which gives 
    \begin{equation} \label{eq:m-h-k-l}
        \begin{split}
            &L_k = (\Bar{A} P^f_{k-1, l} \Bar{C}^\top + \Bar{G} \Tilde{Q} \Bar{M}^\top) (\Bar{C} P^f_{k-1, l} \Bar{C}^\top + \Bar{M} \Tilde{Q} \Bar{M}^\top + R)^{-1},\\
            &m^f_{k, l} = m_{k, l}^- + L_k (y_{k} - \Bar{C} m^f_{k-1, l} - \Bar{D} \, \Bar{u}_k ),\\
            &P^f_{k,l} = P_{k, l}^- - L_k (\Bar{A} P^f_{k-1, l} \Bar{C}^\top + \Bar{G} \Tilde{Q} \Bar{M}^\top)^\top.
        \end{split}
    \end{equation}

    \subsection{Filtering for fast-rate states}
    To find the filtering results for fast-rate states, we first reformulate the dynamic model in Equation~\eqref{eq:x-k-i} in a batch parametrization which stacks all the states on the interval:
    \begin{equation}  \label{eq:fast-dynamic}
                \mathcal{X}_{k, 1:l} = \mathcal{A} x_{k-1, l} + \mathcal{B} \bar{u}_k + \mathcal{G} \bar{w}_k,
    \end{equation}
    where $ \mathcal{X}_{k, 1:l} = \begin{bmatrix}
                x_{k, 1} & x_{k, 2} & \cdots & x_{k, l}
            \end{bmatrix}^\top$ and
    \begin{equation} \label{eq:params-fast}
        \begin{split}
        \mathcal{A} &= \begin{bmatrix}
                A^\top & [A^2]^\top & \cdots & [A^{l}]^\top
            \end{bmatrix}^\top \in \mathbb{R}^{ln_x \times n_x},\\
            \mathcal{G} &= \begin{bmatrix}
                I & 0 & \cdots & 0\\
                A & I & \cdots & 0\\
                \vdots& & \ddots & \\
                A^{l-1} & A^{l-2} & \cdots & I
            \end{bmatrix} \in \mathbb{R}^{ln_x \times l n_x},\\
            \mathcal{B} &= \begin{bmatrix}
                B & 0 & \cdots & 0\\
                AB & B & \cdots & 0\\
                \vdots& & \ddots & \\
                A^{l-1}B & A^{l-2}B & \cdots & B
            \end{bmatrix}\in \mathbb{R}^{ln_x \times l n_u}.
        \end{split}
    \end{equation}
    
    \textbf{Prediction of the fast-rate states, that is, $p(\mathcal{X}_{k, 1:l} \mid y_{1:k-1}) \sim \mathcal{N}(\mathcal{X}_{k, 1:l}; \mathcal{M}_{k, 1:l}^-, \mathcal{P}_{k, 1:l}^-)$} then follows using \eqref{eq:fast-dynamic}:
     \begin{equation} \label{eq:predict-fast}
         \begin{split}
             \mathcal{M}_{k, 1:l}^- &= \mathcal{A} m^f_{k-1, l} + \mathcal{B} \bar{u}_k,\\
             \mathcal{P}_{k, 1:l}^- &= \mathcal{A} \, P^f_{k-1, l} {\mathcal{A} }^\top + \mathcal{G} \, \Tilde{Q} \,  \mathcal{G}^\top.
         \end{split}
     \end{equation}
     
     \textbf{Update of the fast-rate states, that is, $p(\mathcal{X}_{k, 1:l} \mid y_{1:k}) \sim \mathcal{N}(\mathcal{X}_{k, 1:l}; \mathcal{M}^f_{k, 1:l}, \mathcal{P}^f_{k, 1:l})$} can be obtained by noting that given Equation~\eqref{eq:fast-dynamic}, the measurement model in Equation~\eqref{eq:y-k-l} can be rewritten as:
     \begin{equation} \label{eq:y-k-l-2-H}
        y_{k} = \mathcal{H} \mathcal{X}_k + v_{k}, \quad \mathcal{H} = \frac{1}{l}\begin{bmatrix}
                C & \cdots & C 
            \end{bmatrix} \in \mathbb{R}^{n_y \times l n_x}.
    \end{equation}
     Using Equations~\eqref{eq:fast-dynamic} and~\eqref{eq:y-k-l-2-H}, the required update becomes:
     \begin{equation} \label{eq:update-fast}
         \begin{split}
             \mathcal{L}_k &= \mathcal{P}_{k, 1:l}^- \mathcal{H}^\top (\mathcal{H} \mathcal{P}_{k, 1:l}^-  \mathcal{H}^\top + R)^{-1},\\ 
             \mathcal{M}^f_{k, 1:l} &= \mathcal{M}_{k, 1:l}^- + \mathcal{L}_k (y_k - \mathcal{H} \mathcal{M}_{k, 1:l}^-),\\
              \mathcal{P}^f_{k, 1:l} &= \mathcal{P}_{k, 1:l}^- - \mathcal{L}_k \mathcal{H} \mathcal{P}_{k, 1:l}^-.
         \end{split}
     \end{equation}
    \begin{remark} \label{remark:fast-par-filter}
        After obtaining the filtering results for slow-rate states using Equation~\eqref{eq:m-h-k-l}, the filtering estimates for fast-rate states can be computed in parallel over the intervals using Equation~\eqref{eq:update-fast}.
    \end{remark} 
    \begin{remark} \label{remark:cross-covariances}
        The covariances $\mathcal{P}^f_{k, 1:l} \in \mathbb{R}^{n_x \times n_x}$ in Equation~\eqref{eq:update-fast} include all cross-covariances between the fast-rate states within each interval, that is, $\mathcal{P}^f_{k, i, j}$ for $i = \{1, \ldots, l\}$ and $j = \{1, \ldots, l\}$. The filtering covariances can be obtained by extracting $\mathcal{P}^f_{k, i, i}$. However, in the smoother, we also need other covariance blocks than these. It is also possible to reduce the complexity of \eqref{eq:update-fast} by only computing the required parts of the cross-covariance, but the parallelization properties of the computation remain the same.
    \end{remark} 
    \subsection{The proposed sequential smoothing method for the slow and fast-rate states}
    We aim to derive smoothing solutions for all the states, that is, $\big(p(\mathcal{X}_{k, 1:l} \mid y_{1:N}) = \mathcal{N}(\mathcal{X}_{k, 1:l}; \mathcal{M}^s_{k, 1:l}, \mathcal{P}^s_{k, 1:l})\big)|_{k = 1}^N$. To begin, let us first see how the smoothing solution is derived in the conventional RTS smoother.    
    In our notation, the standard RTS smoother corresponds to the case when $l=1$ and is based on the observation that $x_{k, l}$ is independent of $y_{k+1: N}$ given $x_{k + 1, l}$ [Theorem 12.2, Equation 12.9]\cite{sarkka2023bayesian} as follows:
    \begin{equation} \label{eq:smoothing-markov}
        p(x_{k, l} \mid x_{k+1, l}, y_{1:N}) = p(x_{k, l} \mid x_{k+1, l}, y_{1:k}).
    \end{equation}
    However, in SRTM systems where $l > 1$, Equation~\eqref{eq:smoothing-markov} does not hold. Instead, for $i =\{1, \ldots, l\}$ we have
     \begin{equation} \label{eq:srtm-smoothing-markov}
        p(x_{k, i} \mid x_{k+1, 1}, y_{1:N}) = p(x_{k, i} \mid x_{k+1, 1}, y_{1:k}).
    \end{equation}
    Equation~\eqref{eq:srtm-smoothing-markov} says that, given the first state of the next interval, all the states in the current interval are independent of $y_{k+1:N}$. Based on this property, we will construct a model that defines the transition from all fast-rate states of each interval to the first state of the next interval as follows:
    \begin{equation} \label{eq:smoothing-dynamic}
            x_{k+1, 1} = \hat{\mathcal{A}} \mathcal{X}_{k, 1:l} + B {u}_{k, l} + w_{k, l},
    \end{equation}
    where $\hat{\mathcal{A}} = \begin{bmatrix}
        0 & \cdots & 0 & A
    \end{bmatrix} \in \mathbb{R}^{n_x \times lnx}$. Given the filtering distribution of states $\mathcal{X}_{k, 1:l}$ and considering the joint distribution of $x_{k+1, 1}$ and $\mathcal{X}_{k, 1:l}$, we apply a similar procedure as in  [Theorem 12.2]\cite{sarkka2023bayesian} to obtain the smoothing results as:
    \begin{equation} \label{eq:seq-smoothing-all}
        \begin{split}    
          m^-_{k+1,1} &= \hat{\mathcal{A}} \mathcal{M}^f_{k, 1:l} + B u_{k, l}, \\
          P^-_{k+1,1} &= \hat{\mathcal{A}} \mathcal{P}^f_{k, 1:l} \hat{\mathcal{A}}^\top + Q, \\
          G_{k, 1:l} &= \mathcal{P}^f_{k, 1:l} \hat{\mathcal{A}}^\top [P^-_{k+1,1}]^{-1}, \\
          \mathcal{M}^s_{k, 1:l} &= \mathcal{M}^f_{k, 1:l} + G_{k, 1:l} (m^s_{k+1, 1} - m^-_{k+1,1}), \\
          \mathcal{P}^s_{k, 1:l} &= \mathcal{P}^f_{k, 1:l} + G_{k, 1:l} [P^s_{k+1, 1} - P^-_{k+1,1}] G_{k, 1:l}^\top.
        \end{split}    
    \end{equation}
    From Equation~\eqref{eq:seq-smoothing-all} we can see that the smoothing results for all states can be obtained from the smoothing results of the state $x_{k, 1}|_{k = 1}^N$. Thus, by picking suitable blocks from the equations, we can formulate a slow-rate smoother, where the slow-rate states are now $x_{k, 1}|_{k = 1}^N$: 
    \begin{equation} \label{eq:seq-smoothing-1}
        \begin{split}    
          G_{k, 1} &= \mathcal{P}^f_{k, 1, l} A^\top [P^-_{k+1,1}]^{-1}, \\
          m^s_{k, 1} &= m^f_{k, 1} + G_{k, 1} (m^s_{k+1, 1} - m^-_{k+1,1}), \\
          P^s_{k, 1} &= P^f_{k, 1} + G_{k, 1} [P^s_{k+1, 1} - P^-_{k+1,1}] G_{k, 1}^\top.
        \end{split}    
    \end{equation}
    \begin{remark} \label{remark:parallel-fast-smoothing}
        After obtaining the smoothing results for slow-rate states $x_{k, 1}|_{k=1}^N$, the smoothing estimates for other states can be computed in parallel using Equation~\eqref{eq:seq-smoothing-all}.
    \end{remark}
    \begin{remark} \label{remark:pf-k-1-l}
        In Equation~\eqref{eq:seq-smoothing-1}, $\mathcal{P}^f_{k, 1, l}$ is the filtering cross-covariance between the first and the last state in the $k$th interval, see Remark~\ref{remark:cross-covariances}.
    \end{remark}
%
\section{The proposed parallel method}
    \label{sec:proposed-method}
     In this section, our goal is to derive parallelizable formulations for the system with integrated measurements. Specifically, we are interested in computing parallel formulations for all slow-rate states with $O(\log N)$ time complexity. 
    \paragraph*{\textbf{Parallel IMKF}} First, we will derive parallel filter results for slow-rate states $x_{k, l}|_{k=1}^N$. As discussed in Section~\ref{sec:background-parallel}, we need to specify the elements $a_k$ and the binary associative operator, $\otimes$. In the original formulation given in Equation~\eqref{a_k_filter-app}, the conditional density $f_k$ and the likelihood $g_k$ depend on $x_{k-1}$, the state at the previous time step. However, in our scenario with integrated measurements, we determine these densities as functions of $x_{k-1, l}$, representing the last state in the previous time interval as follows:
    \begin{equation}
        \begin{split}
            f_k (x_{k, l} \mid x_{k-1, l}) &= p(x_{k, l} \mid y_{k}, x_{k-1, l}),\\
            g_k(x_{k-1, l}) &= p(y_{k} \mid x_{k-1, l}).
        \end{split}
    \end{equation}
    Densities $p(x_{k, l} \mid y_{k}, x_{k-1, l})$ and $p(y_{k} \mid x_{k-1, l})$ can be found using Equations~\eqref{eq:h-k-l-2} and \eqref{eq:y-k-l-2}. By applying the Kalman filter update with measurement $y_{k}$, applied to the density $p(x_{k, l} \mid x_{k-1, l})$ and matching the terms we have (cf. \cite[lemma 7]{sarkka2020temporal}):
    \begin{equation}
        \begin{split}
            f_k (x_{k, l} \mid x_{k-1, l}) &= p(x_{k, l} \mid y_{k}, x_{k-1, l})\\
            &= N(x_{k, l} ; F_k x_{k-1, l} + d_k, D_k),
        \end{split}
    \end{equation}
    where by defining $R_x = \Bar{M} \Tilde{Q} \Bar{M}^\top + R$, for $k > 1$ we have
    \begin{equation} \label{eq:a-k-f}
        \begin{split}
            K_k &= \Bar{G} \Tilde{Q} \Bar{M}^\top R_x^{-1}, \quad F_k = \Bar{A} - K_k \Bar{C},\\
            d_k &= K_k y_{k} + (\Bar{B} - K_k \Bar{D}) \, \Bar{u}_k,\\
            D_k &= \Bar{Q} - K_k \Bar{M} \Tilde{Q} \Bar{G}^\top,
        \end{split}
    \end{equation}
    and for $k=1$ by using the first step of Kalman filter, that is using Equations~\eqref{eq:prediction} and \eqref{eq:m-h-k-l} with $k=1$, and matching the terms we get $F_1=0$, $d_1 = m_{1, l}^f$, and $D_1 = P_{1, l}^f.$
    
    In order to find $g_k(x_{k-1, l}) = p(y_{k} \mid x_{k-1, l})$, we can use Equation~\eqref{eq:y-k-l-2} and information form of Kalman filter to find $p(y_{k} \mid x_{k-1, l}) \propto N_I(x_{k-1, l}; \eta_k, J_k)$ \cite[lemma 7]{sarkka2020temporal}:
    \begin{equation}
        \begin{split}
            \eta_k & = \Bar{C}^T  R_x^{-1} (y_{k} - \Bar{D} \, \Bar{u}_k),\quad J_k = \Bar{C}^T  R_x^{-1} \Bar{C} .
        \end{split}
    \end{equation}
    Now similarly to \cite{sarkka2020temporal}, given $a_i$ and $a_j$ with the  parameters $a_k=(F_k,b_k,D_k,\eta_k,J_k)$, the binary associative operator $a_i \otimes a_j = a_{ij}$ can be formed as follows (cf. \cite[lemma 8]{sarkka2020temporal}):
    \begin{equation} \label{eq:op-filter}
    \begin{split}
    F_{ij} &= F_j (I_{n_x} + D_i J_j)^{-1} F_i,\\
    d_{ij} &= F_j (I_{n_x} + D_i J_j)^{-1} (d_i + D_i \eta_j) + d_j,\\
    D_{ij} &= F_j (I_{n_x} + D_i J_j)^{-1} D_i F_j^\top + D_j,\\
    \eta_{ij} &= F_i^\top (I_{n_x} + J_j D_i)^{-1} (\eta_j - J_j d_i) + \eta_i,\\
    J_{ij} &= F_i^\top (I_{n_x} + J_j D_i)^{-1} J_j F_i + J_i.
    \end{split}
    \end{equation}
    The Kalman filter means and covariances for $x_{k, l}|_{k=1}^N$ can be extracted from the associative scan results as $d_k$ and $D_k$.
    \paragraph*{\textbf{Parallel IMS}}
    We now aim to derive a parallel integrated measurement smoother (IMS) for states $x_{k, 1}|_{k=1}^N$. Similar to the filtering part, we need to specify the element $a_k$ and the binary associative operator $\otimes$. Using the dynamic model in Equation~\eqref{eq:smoothing-dynamic} and using the same procedure as in \cite[lemma 9]{sarkka2020temporal}, we obtain:
    \begin{equation}
        \begin{split}
            a_k(x_{k, 1} \mid x_{k+1, 1}) &= p(x_{k, 1} \mid y_{1:k}, x_{k+1, 1}) \\
            &= \mathcal{N}(x_{k, 1}; E_k x_{k+1, 1} + g_{k}, S_k),
            \end{split}
    \end{equation}
    where for $k < N$, $ E_k = P^f_{k, 1, l} A^\top (A P^f_{k, l} A^\top + Q)^{-1}$ and
    \begin{equation} \label{eq:parallel-ak-smoothing}
        \begin{split}
            g_k &=  m^f_{k, 1} - E_k (A m^f_{k, l} + B \Bar{u}_{k, l}), \\
            S_k &= P^f_{k, 1} - E_k A P^f_{k, 1, l},
        \end{split}
    \end{equation}
    and for $k = N$ we have $E_N = 0$, $g_N = m^f_{N, 1}$, and $S_N = P^f_{N, 1}$.
    Now, given $a_i$ and $a_j$,  the binary associative operator defined by $a_i \otimes a_j = a_{ij}$ becomes (cf. \cite[lemma 10]{sarkka2020temporal}): 
    \begin{equation} \label{eq:op-smoother}
        \begin{split}
            E_{ij} &= E_i E_j,\, g_{ij} = E_i g_j + g_i,\, S_{ij} = E_i S_j E_i^\top + S_i.
        \end{split}
    \end{equation}
    The RTS smoother's means and covariances for  $x_{k, 1}|_{k=1}^N$ can be extracted from the associative scan results as $g_k$ and $S_k$. 

    As per Remarks \ref{remark:fast-par-filter} and \ref{remark:parallel-fast-smoothing}, given the slow-rate results, the fast-rate filter and smoother results can be computed fully in parallel using \eqref{eq:update-fast} and \eqref{eq:seq-smoothing-all}, respectively. Thus the resulting complexity of the full algorithm is $O(\log N)$.
    
\section{Experimental results}
\label{sec:experiment}
This section empirically evaluates the computational efficiency of our proposed method using a continuously stirred tank reactor (CSTR) case study \cite{gopalakrishnan2011incorporating}. The setup employs an NVIDIA\textsuperscript{\textregistered} GeForce RTX 3080 Ti (12 GB, 10240 CUDA cores) and the Python library JAX~\cite{jax2018github} for parallel processing.

\begin{figure}[tbh]
    \centering
    \resizebox{0.8\columnwidth}{!}{\begin{tikzpicture}
\begin{axis}[
    legend style={
        nodes={scale=0.75, transform shape},
        at={(0, 1)},
        anchor=north west},
    grid=both,
    xmode=log, 
    ymode=log,
    xlabel={$n$ number of time steps},
    ylabel={run time (in seconds)},
    ]
\addplot[gray, line width=1pt] table [x=lengths_space, y=gpu_par_filter_mean_times, col sep=comma]{figures/gpu-final-l16N6000.csv};
\addplot[black, line width=1pt] table [x=lengths_space, y=gpu_par_smooth_mean_times, col sep=comma]{figures/gpu-final-l16N6000.csv};
\addplot[gray, dashed, line width=1pt] table [x=lengths_space, y=gpu_seq_filter_mean_times, col sep=comma]{figures/gpu-final-l16N6000.csv};
\addplot[black, dashed, line width=1pt] table [x=lengths_space, y=gpu_seq_smooth_mean_times, col sep=comma]{figures/gpu-final-l16N6000.csv};
\legend{parallel IMKF, parallel IMS, sequential IMKF, sequential IMS}
\end{axis}
\end{tikzpicture}}
    \caption{GPU run time comparison. \textit{The average RMSE over $100$ runs for $N=200$ is $1.689$ for IMKF and $1.597$ for IMS, respectively.}}
    \label{fig:gpu-run-time}
\end{figure}
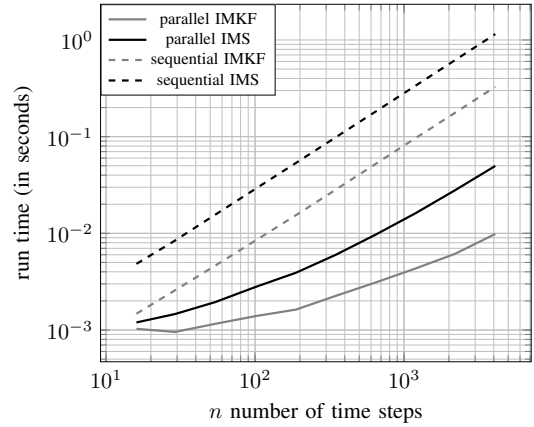

The CSTR model is also applied in \cite{fatehi2017state, yaghoobi2020bayesian, salehi2018state}. The parameters of this model, presented in Equation~\eqref{eq:original-linear-SSM}, are consistent with those in \cite{yaghoobi2020bayesian}.
Measurements are derived by integrating fast-rate samples across an interval of $l = 16$ samples. Observation sets ranging in size from $16$ to $6000$ are examined, and the average run time based on $100$ trials for executing filtering (IMKF) and smoothing (IMS) methods in both sequential and parallel formats is presented in Figure~\ref{fig:gpu-run-time}. As expected, the parallel methods outperform the sequential methods in terms of speed. An open-source implementation is available at \url{https://github.com/Fatemeh-Yaghoobi/Parallel-integrated-method}.
\section{Conclusion}
\label{sec:conclusion}
This paper proposes parallel methods for filtering and smoothing in linear Gaussian systems with integrated measurements. The introduced algorithms reduce the time complexity from linear to logarithmic. Additionally, experiments run on a GPU demonstrate the advantages of these methods compared to traditional sequential approaches.

\bibliographystyle{IEEEbib}
\bibliography{references}

\end{document}